# Cross Event Detection and Topic Evolution Mining in cross events for Man Made Disasters in Social Media Streams

Pramod Bide[1], Sudhir Dhage[2], Mohammed Afaan Ansari[3], Rudresh Veerkhare[4]
[1,2,3,4]Sardar Patel Institute Of Technology, Andheri
Mumbai, Maharashtra, India
[1]pramod_bide@spit.ac.in
[2]sudhir_dhage@spit.ac.in
[3]afaan.ansari@spit.ac.in
[4]rudresh.veerkhare@spit.ac.in

*Abstract*—Social media is widely used to share information globally and it also aids to gain attention from the world. When socially sensitive incidents like rape, human rights march, corruption, political controversy, chemical attacks occur, they gain immense attention from people all over the world, causing microblogging platforms like Twitter to get flooded with tweets related to such events. When an event evolves, many other events of a similar nature have happened in and around the same time frame. These are cross events because they are linked to the nature of the main event. Dissemination of information relating to such cross events helps in engaging the masses to share the varied views that emerge out of the similarities and differences between the events. Cross event detection is critical in determining the nature of events. Cross events have fulcrums points, i.e., topics around which the discussion is focused, as the event evolves which must be considered in topic evolution. We have proposed Cross Event Evolution Detection (CEED) framework which detects cross events that are similar with regards to their temporal nature resulting from main events. Event detection is based on the tweet segmentation using the Wikipedia title database and clustering segments based on a similarity measure. The cross event detection algorithm reveals events that overlap in both time and context to evaluate the effects of these cross events on deliberate/negligent human actions. The topic evolution algorithm puts into perspective the change in topics for an event's lifetime. The experimental results on a real Twitter data set demonstrate the effectiveness and precision of our proposed framework for both cross event detection and topic evolution algorithm during the evolution of cross events.

## I. INTRODUCTION

Social networks are a crucial part of daily life of all users in the world. Thousands of events occur on the social network and users participate in these events by sharing, retweet, tweet, commenting, blogging over social media platforms. Users share their views and opinions leading to continuous evolution of these events. [1]

These concurrently transpiring events over the social network gather research attention in pursuance for detecting evolving events, discovering user interest, detecting influential spreaders, detecting rumours etc. [2]–[11] Twitter is a commonly used microblogging platform where users share short messages of 140 characters called as 'tweets' that are event specific [12]. Social networking sites in latest years playing a vital role in spreading evolving events, including hoax and other misleading messages, rapidly. Social media has become the primary cause of concern in managing man made disasters due to its high tendency of spreading rumors.

Man Made Disaster(MMD) [13] [14] are considered to be more dangerous as compared to natural disasters when it comes to spreading of information about some crisis over social media platform [15]. MMD includes activities like rape, protests, human rights march, terrorists attacks, reservation (Aarakshan). On social media platforms, posts involves criticism, opinions, judgement, and then spreading related but different posts diverting the flow of topic. Such posts create disputes with respect to our main topic. It may happen that when one topic is evolving, it may lead to a series of events, some of which may be cross in nature. Cross events are evolving events which are of similar nature as that of the main event and happen within the same time frame. When a particular event happens, social network is flooded with abundant related tweets. Taking the example of pertaining Rape case #JusticeForAsifa, when this event started evolving few more events came into light, like Candle March, Protests, #JusticeForGeeta and so on. During event evolution multiple subevent follow. Events, #JusticeForAsifa and #JusticeForGeeta are similar in nature and occured in close time span making it imperative to detect cross events with respect to root events. Existing research deals with event detection [16] [8], where as here we are not only considering event detection but also the cross nature of the events. We aim at detecting such two events analysing the effects of such negatively correlated Cross events to understand its contribution to the causes leading to MMD.

When a event is trending i.e., gained a high popularity, then we have observed the manifestation of sub events branching from the root event. During event evolution users shares their thought apropos to the main event due to either their curious nature or to simply share the truth. Users may also end up making contradictory statements or offering condolences. At times, it is simply to gain attention [17]. Correspondingly, an event and it's sub events revolve around varied fulcrum points through the course of its lifetime necessitating the tracking of topic evolution during cross event evolution.

Government officials may use this information to analyze the nature of evolving cross events to understand in what context the events are evolving and easily identify the severity of these events [18].

Given the entire events data that is produced during event evolution, cross event detection and topic evolution algorithms help to track and detect the variations in topics contained in the cross events. Recent research focuses on event detection methods only overlooking cross event detection and topic evolution. On that account, it is important to detect cross events and track evolving topics.

Detecting the evolution of events and their connection with the associated sub-events is essential to codify the fundamental knowledge of how the main event contributes to the evolution of the topic and also to recognize the compelling events. Such events are described as a cluster of countless keywords when identified using most current techniques. The process of recognizing events, categorizing them and summarizing such events is conducted manually. These techniques lacked the effectiveness of identifying important incidents. In addition, current topic models when implemented directly to the microblogging platform consequence in low quality recognition [19] due to information insufficiency issue. To solve this problem, we propose a cross event evolution detection model named CEED. In our proposed model first we identify the topics and detect such events using our text segmentation based Event Detection Algorithm. This algorithm uses tweet segmentation [8] using wikipedia titles database [20] with clustering to find the critical events. All events evolve and it is necessary to understand the flow of events over time. This is addressed in our model through the means of Topic Evolution Algorithm(TEA). The main contributions of this paper are listed as follows:

1) On microblogging platforms like Twitter, it is essential to understand the effects of microblogs will have on the society considering the various events of similar nature occurring in the same time frame. We propose a cross event evolution detection (CEED) model. CEED model considers various parameters as listed in the Table I. Cross events are those two events which occur in the same timespan. Such events are related to the same topic but are not necessarily similar in nature. Taking the example of #JusticeForAsifa and #JusticeForGeeta, even though both events lie under the same area of a topic they are cross in nature.
2) Once the evolution of the topic is analyzed, one should also be able to learn the evolution pattern and predict the possible trailing events. We propose a novel a topic evolution algorithm (TEA). TEA focuses on finding all possible topics in the given Event. TEA algorithm is designed such that it discovers the changes in the progression of corporeality in the event to recognize similarities among the segments that are in focus as the time window shifts through the event lifespan. This algorithm considerably increases the effectiveness of event evolution. For example, the event #JusticeForAsifa further evolved into protests, candle marches, and also inflamed Religious tension in India. Now as we have found out all the topics, we can move towards understanding the evolution pattern. The output of TEA provides an analysis of the evolution of various topics over the lifespan of an Event. This analysis can help us prevent the critical effects of MMD.
3) Experimental results are carried out on the Twitter microblogging platform. The findings after results indicate that the CEED model and TEA algorithm offers comprehensive analytical data on all important evolving events. This shows the effectiveness and precision of our Proposed model for both cross-event detection and topic evolution.

The rest of this paper is organized as follows: First, in Section II, we discuss related research for event detection and event evolution on microblogging platforms. Then, in Section III, we present our CEED model. Finally, we present our experimental results in Section IV. We conclude our research in the last Section V.

## II. RELATED WORK

Topic Modelling is widely used now a days to detect various flooding events on a social network platform. Event detection on a social media using PLSA [21] (Probabilistic Latent Semantic Analysis) has gain attention of many researchers to detect and keep track of all evolving hot events [7]–[10]

PLSA is a statistical technique used for analyzing the co-occurrence of data, in addition to this, it can also be used to find out the hidden variables from some observed variables. Latent Dirichlet allocation (LDA) [22] is a generative statistical model that allows sets of observations to be explained by unobserved groups. It identifies a set of topics belonging to one document. For finding the similarity between topics these models are widely used. However, there have been many state of the art papers discussing the target event detection. But this researches focuses only on topic detection but ignores event evolution with time and how events topics change creating many new events which are the results of the main hot event created earlier. Also, earlier research ignores how such cross-correlated to each other. Existing research does not keep track of evolving topics. [23] [24]

Nikolaos D. Doulamis [9] proposes a fuzzy represented and timely evolved theoretic information matrix for twitter dynamics however the accuracy of this model is not up to mark.

To explore implicit relations within Twitter-based detected events data sets, [4] an online event detection approach is utilized along with word embedding techniques but it fails to detect online cross events, and also accuracy is poor.

Hongyun C. et al [3] proposes a novel event indexing structure known as Multi-layer Inverted List (MIL) which manages the dynamic event databases during updates performed as the event evolves. But this approach incorporates temporal decay which reduces the accuracy and scalability of event detection. VSM (Vector Space models) [25] [26] are also used in many existing research to represent different social events in social media. But, VSM models do not consider other information related to events like key topics and cross nature.

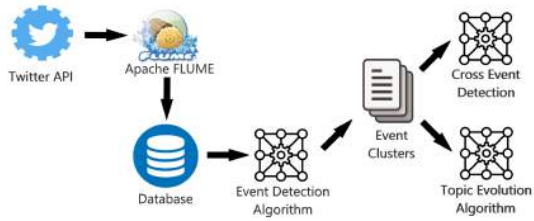

Figure 1. CEED Model system flow

Table I
IMPORTANT PARAMETERS IN CEED

| Sr.no. | Parameter | Description |
|---|---|---|
| 1 | User Count | Count of User related to specific post |
| 2 | Time Similarity | Similarity based on time divergence of events |
| 3 | Contextual Similarity | Similarity of context of events based on the textual data |
| 4 | Post Spread factor | Retweet count for a particular post |

Lei-Lei Shi et al [16] propose a hot event evolution model that considers the short text data in a social network along with the user interest distribution which helps to model the posts and the recommend methods to discover the user interests. Once they identify the users having the great impact it is expected to predict their behavior in near future and how would that affect the event's popularity, but the proposed model was not efficient to do any kind of predictions.

The most related work to ours is the approach proposed in Twevent [8] and SETDWik [20], The concept of text segmentation is used for event detection. As segments are N-grams that contain more relevant information than a single keyword. An external database like Wikipedia is used for efficient segmentation, then ranking segments based on their weights which are calculated using factors like user count and retweet count to extract bursty segments to reduce computational cost and noise present in data. However, these algorithms only deal with event detection and not the cross nature and evolution of the Event.

## III. OUR APPROACH

The Cross Event Evolution Detection (CEED) model, Figure 1 shows us the CEED model which accommodates three modules. Instinctively, first, the event detection algorithm is used to obtain structured event clusters that incorporate key traits of the events contained in the stream of tweets from Twitter API. The clusters formed are used for two major purposes, to detect cross events and to discover the progression of the nature of topics contained in the event. Two separate methodologies are devised for the same. The cross event detection algorithm reveals events that overlap in both time and topics to evaluate the effects of these cross events on deliberate human actions. Topics focused on over the course of an event are subject to change as the event evolves. The topic evolution algorithm puts into perspective the change in topics for an event's lifetime.

### A. Definitions

In this section, we introduce few fundamental concepts used in our paper and the research problem we aim to achieve.

Definition 1 : Tweet
Tweet is a post from a Twitter platform containing the following attributes :
1) tweet_text - Preprocessed text from tweet text fetched from twitter.
2) user_id - twitter id of the user who posted this tweet.
3) retweet_count - number of retweets of the given tweet.
4) time - It's the time of posting of the tweet.

Definition 2 : Segments
Segments are non-overlapping N-grams extracted from processed tweets. The purpose of tweet segmentation is that a phrase contains much more specific information than the unigram or keyword. ex. [say no to war] is much more informative than [say] [no] [to] [war]. Each segment contains the following attributes :
1) tweets - List of Tweets containing a given segment in the current Time Window.
2) frequency - Total number of Tweets containing a given segment in the current Time Window.
3) subwindows- List of Subwindows containing a given segment.
4) newsworthiness- Measure of the importance of a given segment.

Definition 3 : Timewindow and Subwindow
1) Timewindow - Time window is the total period of data collection.
2) Subwindow - Sub window is a unit period in the Time window, ex(day or hour).

Definition 4 : Event
An Event is a cluster of all the tweets and segments with the same context. Each Event contains the following attributes :
1) bursty_segments - List of all Highly weighted segments describing the Event in the current Time Window.
2) tweets - List of all the Tweets in the given Event.
3) time_window -Timewindow containing this Event.
4) subwindows - List of all the Subwindows containing the given Event.
5) event_worthiness - Measure of The importance of a given Event.

Definition 5 : Cross Event
In a stream of events, CE are a pair of overlapping events

⟨ Ei, Ej ⟩ whose parameters (time span,topic) converge in time domain. These events are correlated either positively or negatively.We aim at analysing the effects of such negatively correlated CE events to understand its contribution to the causes leading to MMD.Under MMD we consider the ones caused by one or more recognisable deliberate/negligent human actions.

*B. Event Detection Algorithm (EDA)*

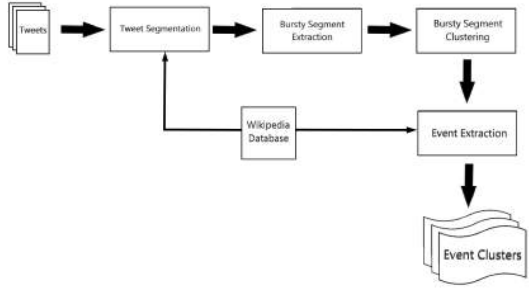

Figure 2. EDA Model

To achieve our goal of identifying the cross events we need to first structurize the tweets into clusters of events. For this purpose we use the Event Detection Algorithm (EDA model), Figure 2 shows us the EDA model. First, the stream of tweets from the Twitter API is retrieved for the time span from February 14th, 2019 to March 2nd, 2019 using Twitter search API. These tweets are then filtered and segmented after that bursty segments are extracted and clustered using Jarvis-Patrick algorithm [27] to get the event clusters.

While filtering the tweet text, firstly tweets are processed to remove non-ASCII characters, links. Hashtags in tweets contain more importance hence more weightage is given to hashtags by increasing their frequency. User names in tweets are replaced with the original names of the users. Capitalization of Hashtags is done since hashtags don't contain white space, ex. #PulwamaTerrorAttack segmented as [pulwama terror attack].

In Twevent (2012) [8] tweets are segmented based on the stickiness score, here we are using an alternative approach based on Wikipedia titles database [20].

Since there might be a very large amount of segments present in real-life data, it's a computationally expensive task to cluster these segments and calculate the similarity between those. Hence to optimize the process and reduce the noise bursty segments are extracted. EDA consists of three algorithms, Firstly Bursty Segment Extractor to extract the most important segments then Bursty Segment Clustering to form the cluster of segments with similar context and finally Event Extraction Algorithm to extract the relevant Events. Output of the Bursty Segment Extractor 1 is provided to Bursty Segment Clustering Algorithm.

**Algorithm 1** Bursty Segment Extractor
**Input:** Segments of Tweets retrieved using Twitter API
**Assumptions:**
$N_t$ - Total number of tweets in given Timewindow
$f_{s,t}$ - Number of tweets containing Segment s in Timewindow t
$rc_{s,t}$ - Sum of retweet count of all tweets containing Segment s
$u_s$ - Count of distinct user related to Segment s
$fc_{s,t}$ - Sum of followers count of all the users realted to Segment s
$P_s$ - Probabity of observing Segment s over the dataset $P_s$ can be considered as Binomial Distribution $B[N_t, P_s]$ with

$$P_s = \frac{f_{s,t}}{N_t} \quad (1)$$
$$E[s|t] = N_t P_s \quad (2)$$
$$\sigma[s|t] = \sqrt{N_t P_s (1 - P_s)} \quad (3)$$

**Calculation:**
1) For each s ϵ Segments
   if $f_{s,t} \geq E[s|t]$ : mark s as Bursty Segment
2) for each Bursty Segment s, Let

$$P_{b(s,t)} = sigmoid[10(\frac{f_{s,t} - [E[s|t] + \sigma[s|t]]}{\sigma[s|t]})] \quad (4)$$

3) weight of segment is given by

$$w_{b(s,t)} = P_{b(s,t)} log(u_{s,t}) log(rc_{s,t}) log(log(fc_{s,t})) \quad (5)$$

4) Sort the Bursty Segments based on weight in descending order and select top $\sqrt{N_t}$ segments for clustering.

**Output:** Bursty Segments describing all Events.

---

**Algorithm 2** Bursty Segment Clustering
**Input:** List of Bursty Segments
**Assumptions:**
$t_m$ - $m^{th}$ Subwindow in Timewindow of total $M$ Subwindows
$f_{t(s,m)}$ - Tweet Frequency of s in subwindow $t_m$
$T_{t(s,m)}$ - Concatenation of all the tweets in given subwindow $t_m$ containing s
$sim(T_1, T_2)$ - TF-IDF similarity between text $T_1$ and $T_2$
**Calculation:**
1) For $s_a$, $s_b$ ϵ Bursty Segments

$$w_{t(s_a,m)} = \frac{f_{t(s_a,m)}}{f_{s,t}} \quad (6)$$
$$w_{t(s_b,m)} = \frac{f_{t(s_b,m)}}{f_{s,t}} \quad (7)$$
$$sim_{t(s_a,s_b)} = \sum_{m=1}^{M} [w_{t(s_a,m)} w_{t(s_b,m)} sim(T_{t(s_a,m)}, T_{t(s_b,m)})] \quad (8)$$

2) Using Jarvis-Patrick clustering Algorithm on similarity matrix obtain the Events Cluster.

**Output:** Event clusters i.e all similar segments grouped into a single event cluster.

In Algorithm 2 Jarvis-Patrick Algorithm [27] is used for clustering. In this, all segments are considered as nodes and initially, all nodes are disconnected. An edge is added between segments $s_a$ and $s_b$ if k-Nearest neighbors of $s_a$ contains $s_b$ and vice versa. After adding all possible edges, all the connected components of the graph are considered as candidate event clusters. Those segments that do not have any edges are discarded from further processing.

Output of the Algorithm 2 is well clustered events However some events are not relevant, to filter such events out event-worthiness is calculated with the help of Wikipedia Database.

**Algorithm 3** Event Extraction
**Input :** Clusters of Events
**Assumptions :**
$e_s$ - set of segments in Event Cluster
$e_g$ - set of similarity values between segments in Event Cluster
$Q(s)$ - probability of $s$ appearing as anchor text in Wikipedia article
$\mu(s)$ - newsworthiness of $s$

$$\mu(s) = \begin{cases} e^{Q(s)} & \text{if } s \text{ is a word} \\ max_{l \epsilon s} e^{Q(s)} - 1 & \text{otherwise} \end{cases} \quad (9)$$

$\mu(e)$ - eventworthiness of Event cluster $e$
$\tau$ - Threshold
**Calculation:**
1) for each $e \, \epsilon$ Event Clusters

$$\mu(e) = \frac{\sum_{s \epsilon e_s} \mu(s)}{|e_s|} \times \frac{\sum_{g \epsilon e_g} sim(g)}{|e_s|} \quad (10)$$

2) for each $e \, \epsilon$ Event Clusters
if $\frac{\mu_{max}}{\mu(e)} \geq \tau$: select Event Cluster
else discard the Event Cluster

**Output:** Filtered Event Clusters with event-worthiness score.

*C. Cross Event Detection Algorithm (CEDA)*

Cross Event Detection Algorithm (CEDA Model) can be observed in Figure 3. Considering the amount of tweets generated on the microblogging platform in consideration, Twitter, it becomes imperative to realize the effect that these tweets have on driving the views of the society as a whole. Hence, it becomes necessary to not only detect the nature of events in singularity but also their nature in regards with events of a similar nature occurring in the same time frame. Keeping this essential in mind, the cross event detection algorithm (CEDA) is designed. Refer algorithm 4

From a stream of events the algorithm obtains a pair of events that are overlapping in nature. A pair of events $\langle E_i, E_j \rangle$ classify as cross events when their topic exhibits a relation between each other in time domain.

For two events to be cross in nature there are two major factors to consider, first is Contextual Similarity between the Events and second is Time Similarity between two Events. Contextual Similarity is calculated based on the similarity of clustered Segments and Tweets in Events. Time similarity is calculated based on the overlapping nature of Events in terms of time.

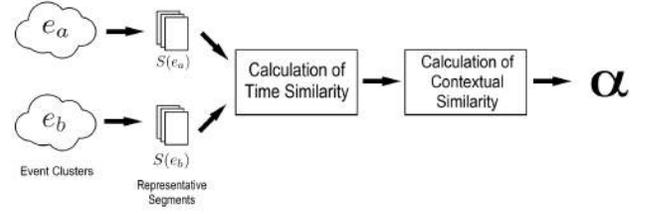

Figure 3. Cross Event Detection Algorithm

In CEDA(Algorithm 4) Representative Segments $S(e)$ are extracted from Event $e$ by sorting all segments present in Event Cluster $e$ in descending order of there probability of occurrence and selecting $\sqrt{N}$ top segments where $N$ is the total count of segments in Event Cluster. After this, Time similarity for each pair of segment $t(s_a, s_b)$ in calculated, using $t(s_a, s_b)$ we calculate the Time similarity between events $t(e_a, e_b)$. To calculate Contextual similarity TF-IDF similarity between concatenation of all segmented tweets in events is calculated. Then the product of both the similarities gives the cross event factor, we're using the hyperbolic tangent function to get a normalized value between 0 and 1.

**Algorithm 4** Cross Event Detection Algorithm
**Input:** Two events clusters $e_a$ and $e_b$
**Assumptions :**
$S(e)$ - Representative Segments from event $e$
$T_s(e)$ - Concatenation of all segmented tweets in event $e$
$t(s_a, s_b)$ - Time Similarity between two segments $s_a$ and $s_b$
$t(e_a, e_b)$ - Time Similarity between two events $e_a$ and $e_b$
$C_s(e)$ - Count of all representative segments in event $e$
$w_t(s, m)$ - Fraction of frequency of $s$ in Subwindow $t_m$ as given in (7)
$\alpha(e_a, e_b)$ - Cross Event factor
**Calculation:**
1) For each $s_a \, \epsilon \, S(e_a)$ and $s_b \, \epsilon \, S(e_b)$ Calculate $t(s_a, s_b)$.

$$t(s_a, s_b) = \sum_{m=1}^{M} [w_{t(s_a, m)} w_{t(s_b, m)}] \quad (11)$$

2) For each $s_a \, \epsilon \, S(e_a)$ and $s_b \, \epsilon \, S(e_b)$

$$t(e_a, e_b) = \frac{\sum_{s_a \epsilon S(e_a)} [\sum_{s_b \epsilon S(e_b)} t(s_a, s_b)]}{C_s(e_a) \times C_s(e_b)} \quad (12)$$

3) Calculating Cross similarity factor

$$\alpha(e_a, e_b) = tanh(t(e_a, e_b) \times sim(T_s(e_a), T_s(e_b))) \quad (13)$$

**Output:** Degree to which events are cross in nature

## D. Topic Evolution Algorithm (TEA)

Deriving from the event detection algorithm, we obtain Event Clusters. Consequently, each event cluster is a segregated group of segments with similar context and traits, hence revolving around a set of closely related topics in the course of the lifetime of the event. The context of the focus of event topics is liable to change in due course of the event with respect to time. It can be simply stated that if a particular topic is key at the start of the event, it may or may not continue to be key by the closure of the said event.

Taking the example of the BhimaKoregaon event, which started with a topic like protests, going on to focus on bandh and reservation/aarakshan where the context transformed from activism to politics. Given this, the Topic evolution algorithm (TEA) focuses on discovering the changes in the progression of corporeality contained in the event cluster.

In Topic Evolution Algorithm, firstly we filter the segments given by an Event cluster by sorting the segments in descending order by there occurrence probability then filter out the segments with freq more than $N/2$ and select first $\sqrt{N}$ segments where $N$ is the total number of segments. After this cluster the topic with Jarvis-Patrick Algorithm to get the Topic Clusters. In the final step, we calculate the popularity if a Topic in each Subwindow by summing the subwindow wise probability of each segment in the Topic cluster. The output of the algorithm contains the trend of each topic over time of the given Event. Refer Algorithms 5

---

**Algorithm 5** Topic Evolution Algorithm

**Input :** Event cluster $e$
**Assumptions :**
$S_f(e)$ - filtered segments from event cluster $e$
$sim_t(s_a, s_b)$ - segment similarity as given in equation 8
$P_t(s, m)$ - Probability of segment $s$ occurring in $m^{th}$ Subwindow

$$P_t(s,m) = \frac{\text{frequency of } s \text{ in } m^{th} \text{ subwindow}}{\text{Total count of segments in } m^{th} \text{ subwindow}} \quad (14)$$

**Calculation :**
1) For $s_a$, $s_b$ $\epsilon$ $S_f(e)$
   Calculate $sim_t(s_a, s_b)$
2) Using Jarvis-Patrick Algorithm for clustering Cluster the segments into Topic Cluster using similarity score calculated above.
3) For $T$ $\epsilon$ Topic Cluster,
   For each $m$ $\epsilon$ $\{1, 2...M\}$ Calculate
   $$T(m) = \sum_{s \epsilon S_T} P_t(s,m) \quad (15)$$
   Where $S_T$ are segments in Topic Cluster $T$

**Output:** Topic Clusters with degree of popularity of each topic in every Subwindow.

---

## IV. EXPERIMENTATION RESULTS

In this section, we communique the results of an extensive study conducted on a vast real-life tweet dataset.As we know, our proposed CEED model can detect various events and identify cross events amongst them.The experiments are designed to evaluate the accuracy and efficiency of CEED. The results show events detected, identified Cross Events and the effectiveness in mapping evolution of the various detected Cross Events.

And the rest of the section talks about, our gathered collection of data,experimental setup and analysis, the baseline approach and the model evaluation.

### A. Dataset

Our dataset is generated from Twitter [12] via Twitter API. This dataset consists of 2,78,817 tweets posted from February 15th, 2019 to March 20th, 2019 using Twitter search API. Here everything will get in JSON format and this is stored in the HDFS that has given the location where to save all the data that comes from Twitter. After running the Flume, the Twitter data will automatically save into HDFS.

Every tweet in our dataset undergoes preprocessing which removes non-ASCII characters and stop words, stemming, and obtains the appropriate segments. Further, we consider only those tweets which are unique (do not count the tweet which is just a retweet). Again for the users, we count the ones who have either posted the tweet or retweeted on the tweet. We get a total of 45,948 high-quality tweets.

### B. Evaluation Phases

For our experiments we employed the dataset containing microblogs on the events that occurred in India,i.e. The Pulwama Attack. The events present on Twitter platform are highly heterogeneous. Our work is to first cluster all the events together, find the similarity between them and keep a track on how they evolve in future.This information then helps us to predict event evolution for another topic under same category.

*1) **Event Detection Algorithm**:* Firstly the segmentation is carried out using Wikipedia titles database then the segmented tweets are clustered and each event cluster is described as the cluster of segments Table II shows the events and the bursty segments related to each event. Label is the label given to the Event to identify it based on the set of bursty segments.

There were several parameters that affect the performance of the model like hashtag weight $H$, number of neighbors $k$ while clustering, and threshold $\tau$. We set $H$ = 3, $k$ = 4 neighbors and $\tau$ = 4 in our work.

Table II
STRUCTURIZED EVENT CLUSTERS

| Event | Label | Bursty Segments |
|---|---|---|
| E1 | say no to war | [ say no to war, bring back abhinandan , saynotowar, pakistan strikes back, pakistan army zindabad, say noto war, pakistan zindabad, peace war ] |
| E2 | blood donation by dss | [ blood donation by dss, dera sacha sauda, blood donation, drgurmeet ram rahim, donate blood, dr singh, gurmeet ram rahim singh ] |
| E3 | abhinandan returns | [ abhinandan returns, welcome home abhinandan, welcome home abhinanadan, abhinanadan varthaman, welcome back, real hero ] |
| E4 | pulwana attack | [ pulwama, crpf, pulwama attack, kashmir terror attack, crpf jawans, rip brave hearts ] |
| E5 | me too | [ me too, metoo, times up, surviving r kelly, mute r kelly, r kelly, suriving r kelly ] |

Given event clusters are filtered based on their event worthiness score $\mu(e)$ as given in Table III. Which is calculated based on Algorithm 3.

Table III
EVENT WORTHINESS OF EVENTS

| Event | Score |
|---|---|
| E1 | 1.7614331627247717 |
| E2 | 1.6606617003112545 |
| E3 | 1.5248289851004615 |
| E4 | 1.412936448555728 |
| E5 | 0.693081912381193 |

After the clustering of segments, we classified tweets according to the Events by classifying tweet based on its segments and the Event wise count of tweets is shown in Figure 5.

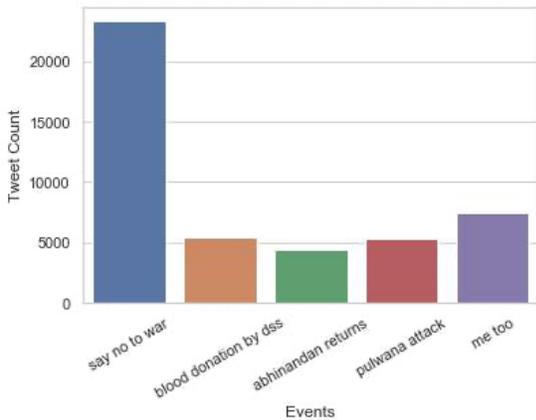

Figure 4. Event wise Tweet count

*2) Cross Event Detection Algorithm:*

*a) Cross Event Detection Matrix:* It defines the degree to which all events are cross in nature. Using the CEDA algorithm the factor of cross incidents is calculated. Table IV demonstrates us how events are cross in nature.

Table demonstrates that E1 and E3 are cross in nature as events **say no to war** and **bring back abhinandan** was going on in the same context. Similarly, E2 and E4 are cross in nature as event **blood donation** was evolved related to **pulwama attack**. There is small crossness between events E1, E4 and E3, E4 which is because all three events are related to war and india-pakistan.

Table IV
CROSS EVENT MATRIX

| Cross Event Matrix | E1 | E2 | E3 | E4 | E5 |
|---|---|---|---|---|---|
| E1 | 1.0 | 0.07 | 0.95 | 0.43 | 0.10 |
| E2 | 0.07 | 1.0 | 0.02 | 0.89 | 0.01 |
| E3 | 0.95 | 0.02 | 1.0 | 0.41 | 0.06 |
| E4 | 0.43 | 0.89 | 0.41 | 1.0 | 0.05 |
| E5 | 0.10 | 0.01 | 0.06 | 0.05 | 1.0 |

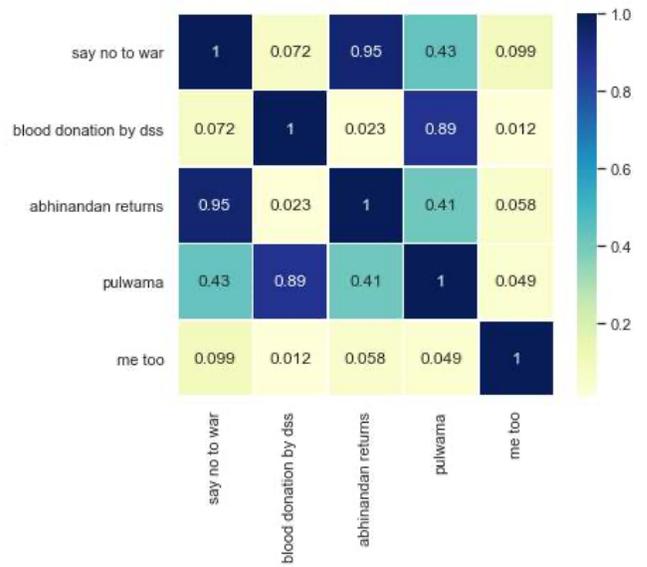

Figure 5. Cross Event Heat Map

*3) Topic Evolution Algorithm:* The topic evolution algorithm is concerned with the pattern of evolution of topics within clustered events. With respect to algorithm 5, topics are identified over the life span of the event and then the weightage of each topic is calculated in every subwindow to get the evolution timeline of the event.

Following figures show the topic evolution timeline of the each event cluster. Each event is divided in to 10 subwindows and topics are detected. The segments related to each topic are given in Table next to graph.

*a) E1 (say no to war)*

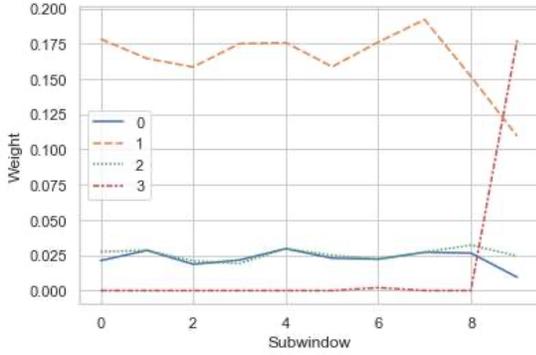

Figure 6. Topic Evolution Timeline in E1

Table V
SEGMENTS DESCRIBING TOPICS IN E1

| Topic No. | Segments |
|---|---|
| 0 | 'peace peace', 'war peace', 'peace war' |
| 1 | 'bring back abhinandan', 'say noto war', 'saynotowar' |
| 2 | 'wing commander', 'jai hind', 'abhinandan varthaman' |
| 3 | 'go back modi', 'nobel peace prize for imran khan', 'pakistan leads with peace' |

*b) E2 (blood donation by dss)*

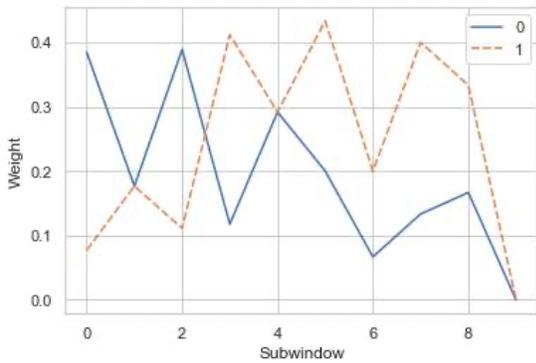

Figure 7. Topic Evolution Timeline in E2

Table VI
SEGMENTS DESCRIBING TOPICS IN E2

| Topic No. | Segments |
|---|---|
| 0 | 'guinness world records', 'world records', 'many records' |
| 1 | 'donating blood', 'life blood', 'true blood', 'always ready' |

*c) E3 (bring back abhinandan)*

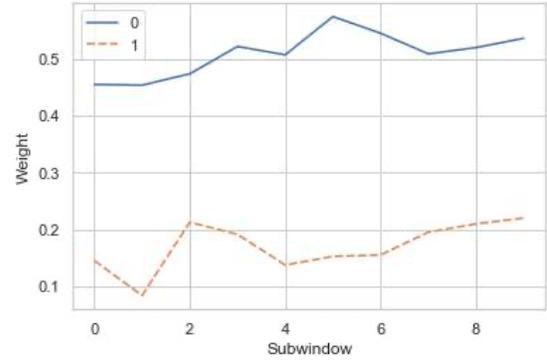

Figure 8. Topic Evolution Timeline in E3

Table VII
SEGMENTS DESCRIBING TOPICS IN E3

| Topic No. | Segments |
|---|---|
| 0 | 'welcome back', 'abhinandan diwas', 'welcome home abhinanadan', 'real hero', 'abhinandan is back', 'welcome home', 'welcome home abhinandan', 'abhinanadan varthaman' |
| 1 | 'india', 'indian air force', 'imran khan', 'say no to war', 'narendra modi', 'pakistan', 'iaf', 'abhinandan' |

*d) E4 (pulwama attack)*

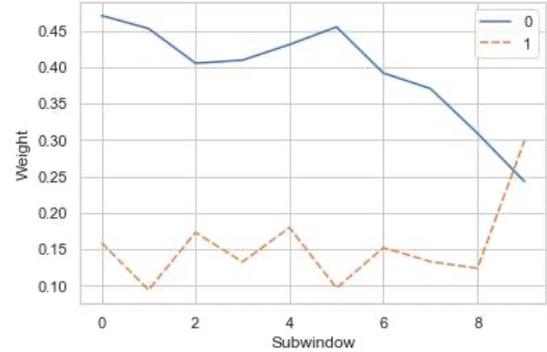

Figure 9. Topic Evolution Timeline in E4

Table VIII
SEGMENTS DESCRIBING TOPICS IN E4

| Topic No. | Segments |
|---|---|
| 0 | 'crpf kashmir attack', 'attack convoy', 'crpf', 'kashmir terror attack', 'pulwana attack', 'crpf jawans', 'rip brave hearts', 'terrorist attack' |
| 1 | 'india', 'kashmir', 'surgical strike', 'imran khan', 'narendra modi', 'pakistan' |

*e) E5 (me too)*

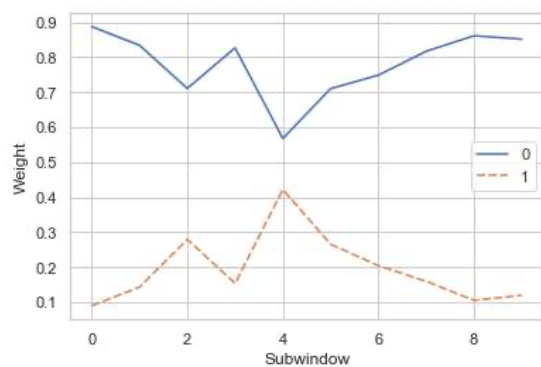

Figure 10. Topic Evolution Timeline in E5

Table IX
SEGMENTS DESCRIBING TOPICS IN E5

| Topic No. | Segments |
|---|---|
| 0 | 'golden globes', 'bryan singer', 'black girls matter', 'goldenglobes', 'surviving kelly', 'red carpet', 'survivingrkelly', 'lifetime', 'surviving r kellly', 'suriving r kelly', 'black women', 'r kelly', 'timesup', 'tarana', 'last year', 'black men', 'ryan seacrest', 'golden globe awards', 'times up', 'muterkelly', 'white women', 'mute r kelly', 'metoo', 'rkelly', 'surviving r kelly' |
| 1 | 'street harassment', 'bernie sanders', 'youtube', 'mission china', 'james barron', 'sexual abuse', 'sexual assault', 'sexual harassment', 'feminism', 'political vigilante', 'new york', 'me too india', 'liked video', 'alok nath', 'anticipatory bail', 'grace notes', 'st patrick' |

From all the above graphs overall life span of the event can be summarized, which helps to understand the evolution pattern of the events and it's useful to predict future behavior.

## V. CONCLUSION

Social media is widely used to share information globally and it also aids to gain attention from the world. However, Discussions in state-of-the-art papers are restricted mostly to event detection and overlook the cross event detection and topic evolution with respect to cross events. This leads to difficulty in tracing the nature of evolving events. To address this problem, We propose the CEED Model. It consists of two main algorithm Cross Event detection Algorithm (CEDA) and Topic Evolution Algorithm (TEA). Cross event detection is critical in determining cross events, which have many fulcrums points, so the CEDA algorithm detects cross events that are similar with regards to their temporal nature resulting from main events. The topic evolution algorithm is concerned with the pattern of evolution of topics within clustered events. Moreover, this approach can further find topics that are shaping the behavior of the event. Finally, the experimental results on the real Twitter dataset demonstrate the efficiency of our proposed model for both cross event detection and topic evolution.

In future work, to understand the impact and evolution of growing events on a larger radii of society, we plan to further predict the behaviors of user community based on the dynamic community detection model . To improve the efficiency of our model we can integrate other social media data e.g Instagram check-in data for accurate predictions.